\documentclass[ twocolumn]{aastex631}

\usepackage{amsmath}
\usepackage{mhchem}
\usepackage{multirow}
\usepackage{boldline}
\usepackage{hhline}
\usepackage{booktabs}
\usepackage{ulem}
\usepackage{graphicx} 
\usepackage{soul}
\usepackage[para]{threeparttable}

\usepackage[caption=false]{subfig}

\shorttitle{AASTeX v6.3.1 Sample article}
\shortauthors{Bovolenta et al.}

\graphicspath{{./}{figures/}}

\begin{document}
\title{Binding Energy Evaluation Platform: A database of quantum chemical binding  energy distributions for the astrochemical community.}

\correspondingauthor{Stefan Vogt-Geisse}
\email{stvogtgeisse@qcmmlab.com}

\author[0000-0001-8908-9109]{Giulia M. Bovolenta}
\affiliation{Departamento de F\'isico-Qu\'imica, Facultad de Ciencias Químicas, Universidad de Concepci\'on, Concepci\'on, Chile}

\author[0000-0002-3102-1774]{Stefan Vogt-Geisse}
\affiliation{Departamento de F\'isico-Qu\'imica, Facultad de Ciencias Químicas, Universidad de Concepci\'on, Concepci\'on, Chile}

\author[0000-0003-2814-6688]{Stefano Bovino}
\affiliation{Departamento de Astronom\'ia, Facultad Ciencias F\'isicas y Matem\'aticas, Universidad de Concepci\'on,
Av. Esteban Iturra s/n Barrio Universitario, Casilla 160, Concepci\'on, Chile}

\author[0000-0002-3019-1077]{Tommaso Grassi} 
\affiliation{Centre for Astrochemical Studies,Max-Planck-Institut f\"ur extraterrestrische Physik,Giessenbachstrasse 1, 85749 Garching bei M\"unchen, Germany}

\begin{abstract}
\noindent The quality of astrochemical models is highly dependent on reliable binding
energy (BE) values that consider the morphological and energetic variety of binding sites on 
the surface of ice-grain mantles. Here, we present the Binding 
Energy Evaluation Platform (BEEP) and database that, using quantum chemical methods,
produces full BE distributions of molecules bound to an amorphous solid water (ASW) surface model. BEEP is highly automatized and allows to sample binding sites on set of water clusters and to compute accurate BEs.
Using our protocol, we computed 21 BE distributions of interstellar molecules and radicals on an amorphized set of 15-18 water clusters of 22 molecules each.
The distributions contain between 225 and 250 unique binding sites.  We apply a Gaussian fit and report the mean and standard deviation for each distribution. 
We compare with existing experimental results and find that the low and high coverage
experimental BEs coincide well with the high BE tail and mean value of our distributions, respectively.  Previously reported single BE theoretical values are broadly in line with ours, even though in some cases significant differences can be appreciated. We show how the 
use of different BE values impact a typical problem in astrophysics, such as the computation of snow lines in protoplanetary discs. BEEP will be publicly released
so that the database can be expanded to other molecules or ice-models 
in a community effort.
\end{abstract}

\section{Introduction} \label{sec:intro}

\noindent In dense interstellar clouds, where the temperature is less than 20 K, interstellar 
dust particles are covered with a layer of ice consisting mostly of \ce{H2O} and, at a
lower proportion,
molecules such as \ce{CO2}, \ce{NH3} and \ce{CH4} \citep[see e.g.][]{boogert_observations_2015}. In these cold
environments, interstellar chemistry can take place on the ice mantles
of interstellar  dust grains \citep[e.g.][]{herbst2009a}. The ice surface is capable of binding different molecules
from the gas phase, thus  facilitating chemical encounters and promoting the formation of new molecular species, that can  be detected once they desorb into the gas phase \citep[e.g.][]{jorgensen2020}. 
In that regard, the binding energy (BE) is a crucial parameter
when modeling gas-grain chemistry in dense clouds, as it determines the desorption rate
of the adsorbed species for  thermal, chemical and photo desorption \citep{minissale2022}.  Having  
knowledge  of the BE of molecules on ice mantles allows astrochemical 
gas-grain models to predict the abundances of molecular and atomic species.\\
The composition, structure and formation of the ice  mantles is still a 
matter of research. However, the broad shape of the water 3.1 $\mu$m O-H stretching band observed in different
dense cloud regions, suggests, upon comparison with experimental results, that the water component of the ice mantles exists in amorphous form, as layers of amorphous solid water (ASW)
\citep{smith1989}. This is important inasmuch the BE depends both on the nature of the adsorbed species and the 
composition and morphology of the ice mantle.\\
BEs on ice surfaces can be determined experimentally, mainly using Temperature
Programmed Desorption (TPD). In TPD experiments, a layer of ASW is build 
through vapor deposition and exposed to the species of interest in a constant
temperature regime. Once the desired level of coverage 
is reached, the temperature is increased and the desorbed molecules are collected
and analyzed by  mass spectrometry.
To date, several TPD experiments have been performed using ASW ice as substrate, ranging from multilayer to 
sub-monolayer regime of adsorbed molecules. One of the first extensive 
TPD studies, done by
\cite{collings2004}, made desorption rate  measurements 
of 16 astrophysically-relevant molecules on an ASW substrate in a monolayer (ML) and multilayer regime. BEs
at 
sub-monolayer deposition  have also been determined using TPD measurements, by inversion of the Polanyi-Wigner
equation, which yields a coverage dependent adsorbate BE. The coverage is usually measured as a fraction of a
ML and ranges from 1 ML to 10$^{-3}$ ML.  Coverage-dependent BE distributions  
have been obtained for a few astrophysically important molecules, such as \ce{N2} \citep{smith2016,he2016a}, \ce{O2} \citep{smith2016,he2016a}, CO \citep{noble2012,smith2016,he2016a}, \ce{CO2} \citep{noble2012,he2016a}, \ce{CH4} \citep{smith2016,he2016a} and \ce{D2} \citep{amiaud2006,he2016a}.
Even though TPD experiments provide valuable BE data, the preparation of the substrate and deposition 
technique can vary among experiments, which makes it difficult to construct a homogeneous database of experimental
BE values. Also, TPD is not suitable to provide BE values for radicals due to the short life-span of these species. 

\noindent On the other hand, BEs can also be determined using a computational approach by means of \textit{ab initio}  quantum chemistry methods and molecular dynamics (MD) simulations. In recent years, there has been important progress in the development of both the construction of ASW models and in the computations of BEs.  
Two types of ASW models have been proposed: using a slab of ASW with periodic boundary conditions, or using
amorphized water clusters. 
In the most complete study thus far, using the former approach,  
\cite{ferrero2020} computed BEs of 21 molecules and atoms. 
Their ASW water slab consisted of 
60 molecules and they computed the BEs for up to 8 binding sites 
per molecule. The cluster approach consists of one or several water clusters 
to simulate parts of the ASW surface. Within the cluster approach, two strategies for computing BE have been proposed. First,
using a large surface of hundreds of water molecules in a QM/MM embedded regime, 
in which the bulk is described with a force field  
and the molecules close to the binding site are computed by means of quantum chemistry methods. 
Using this approach \citet{song2016,song_tunneling_2017} computed BE distributions of HNCO and \ce{H2CO}. More recently \cite{duflot2021} obtained binding energies of 8 different binding sites of several species
(H, C, N, O, NH, OH, \ce{H2}O, \ce{CH3}, \ce{NH3}) using 
a ONIOM QM/QM hybrid method. A similar procedure was used by \cite{sameera2021} to compute 10 binding sites
of the \ce{CH3O} radical.
The other approach to the cluster model
was first introduced by \cite{shimonishi2018}. They used a  set of previously annealed 20- molecule water clusters to represent different regions of an ASW surface. This set of water clusters was sampled with different 
atomic species to compute BEs at  Density Functional Theory (DFT) level of theory, and 
only the highest BE values on each water cluster were reported. 
Based on this set-of-clusters approach, we developed 
a computational procedure to generate BE distributions  and showcased the procedure 
on the HF molecule adsorbed on a set of 22- and 37-molecule clusters, 
considering 255 and 126 unique binding sites, respectively \citep{bovolenta2020a}. 
Recently, \citet{germain2022} computed a BE distribution for \ce{NH3} molecule employing a 
single 200 water icy grain constructed by the semiempirical tight-binding GFN2 method.
Finally, the efforts to obtain an extensive BE catalogue for small molecules on water surfaces
have been limited to DFT computations on small water 
clusters (up to 6 molecules, \citealp{sil2017,das2018}) or interaction with water monomer by 
linear semi-empirical models \citep{Wakelam2017}, which do not capture the complete
statistical nature of the interaction on ASW. However, using a full BE distribution reflects 
a more realistic desorption behaviour for  molecules adsorbed on ASW ice as suggested in \citet{grassi2020a}. Notwithstanding, computing a large set of BEs requires a significant amount of computational resources and data management.

\noindent In this work, we present BEEP, a Binding Energy Evaluation Platform meant to offer a straightforward, highly automated and
easy-to-use interface for the computation and processing of full BE distributions of molecules. To present
the utility of BEEP, we computed BE distributions of 21 astrophysically-relevant 
molecules. The platform is implemented within the QCArchive framework \citep{smith2020a}  
which allows to transform the database in a fully open-source endeavour, from the data generation 
to the final user-query of the BE data.

\section{Computational Details} \label{sec:methodology}

\subsection{Surface Modeling}
\noindent To build an ASW surface serving as an ice mantle model, we adapted the cluster approach, first introduced by \cite{shimonishi2018}.
The initial water cluster, consisting of 22 molecules (\ce{W_{22}}), has been generated by molecular dynamics, using the TIP3P model. We then performed 100 ps of high temperature (300 K) \textit{ab initio} molecular dynamics (AIMD) simulation at BLYP/def2-SVP\citep{becke1988,lee1988, miehlich1989, weigend2005} level of theory, in order to amorphyze the system. We extracted 100 independent structures ($\tau_{correlation}$  $\simeq$ 1 ps) from the resulting trajectory, 
which underwent temperature annealing of 3 ps to reach the target interstellar conditions ($\sim 10$ K).  
We selected the 20 most representative \ce{W_{22}} clusters, grouping the structures according to geometrical criteria (similarity threshold of root-mean-square deviation of atomic
positions (RMSD) $\leq$ 0.40 \text{\AA}).
The surface spanned by these 20 clusters represents our ASW model.\\
The use of clusters of this size allows a good compromise between accuracy and computational time 
and has been validated in our previous work, to which we refer for further details \citep{bovolenta2020a}. 

\subsection{Geometry optimization and binding energy calculation}
\noindent We performed a DFT geometry benchmark on the \ce{W_{1-3}-X} systems, with X being the target molecule and W the water cluster, using DF-CCSD(T)-F12/cc-pVDZ-F12 \citep{bozkaya2017, werner2020, DunningT.H.2001} geometry as a reference (see Appendix \ref{sec:bench_results}, Table \ref{tab:bench}). We also conducted an energy benchmark, using \ce{W_{4}-X} system to compare DFT BE values 
to a CCSD(T)/CBS \citep{klopper1986,feller1992,helgaker1997,karton2006} reference energy (see Appendix, \ref{sec:bench_results}, Table \ref{tab:bench}).
We used BLYP/def2-SVP
as level of theory for the binding site 
sampling procedure by means of the \textsc{Terachem} software \citep{ufimtsev2009,titov2013}, to take advantage of
the efficient GPU acceleration. All high level 
DFT optimizations were preformed together with a def2-TZVP basis set. We also computed the Hessian matrix for selected structures at the equilibrium geometry
to obtain the Zero-Point Vibrational Energy (ZPVE) contribution  to the BE, computed at the same level of theory as the geometry optimization.\\
The BE has been calculated as 

\begin{equation}
    \Delta E_b = \Delta E_{CP} + \Delta_{ZPVE} \,,
\end{equation}
\noindent with $\Delta E_{CP}$ being the binding electronic energy corrected for the basis set superposition error (BSSE) and $\Delta_{ZPVE}$ the ZPVE correction for the BE.
See the Appendix \ref{sec:BE_stoich},\ref{sec:ZPVE} for more details.
We consider the BE as a positive quantity, according to convention.
For the single point computations at 
DFT level of theory, we employed a def2-TZVP basis 
set. All high level optimization and energy computations were performed using \textsc{Psi4} \citep{parrish2017}.

\subsection{QCArchive Framework}\label{sec:qcarchive}
\noindent Quantum chemistry data has been traditionally generated through user-defined
individual input files, which are processed by a specific software that
stores the results of the computation in output files. These outputs are then parsed 
either by hand or using custom scripts. This approach has serious
limitations when attempting to compute a large volume of 
data as it is error-prone and the results are difficult to reproduce, since parsing scripts and 
output files are usually not available.
To overcome these limitations, we build the 
BEEP platform within the Python-based QCArchive framework. The details about the different components 
of the QCArchive infrastructure have been described elsewhere \citep{smith2021}.
The core component of BEEP is a central server to which computation results 
are added in the form of JSON (JavaScript Object Notation) objects  that contain 
the same level of information as a traditional  output file. 
The access to this database, where the user can query existing data and submit additional computations, is controlled by a standard username/password system.  Moreover, several
data objects can be defined to generate and sort the data.
These collections (called \textit{Datasets}) make it possible to 
extend a procedure, such as a geometry optimization or a BE computation,
to a large number of objects in a single operation. 
Finally, the generated values can be easily accessed
from the stored collections. 

\section{Results}\label{sec:results}

\begin{figure*}[ht]
    \centering
    \includegraphics[angle=0, width = 0.9\linewidth]{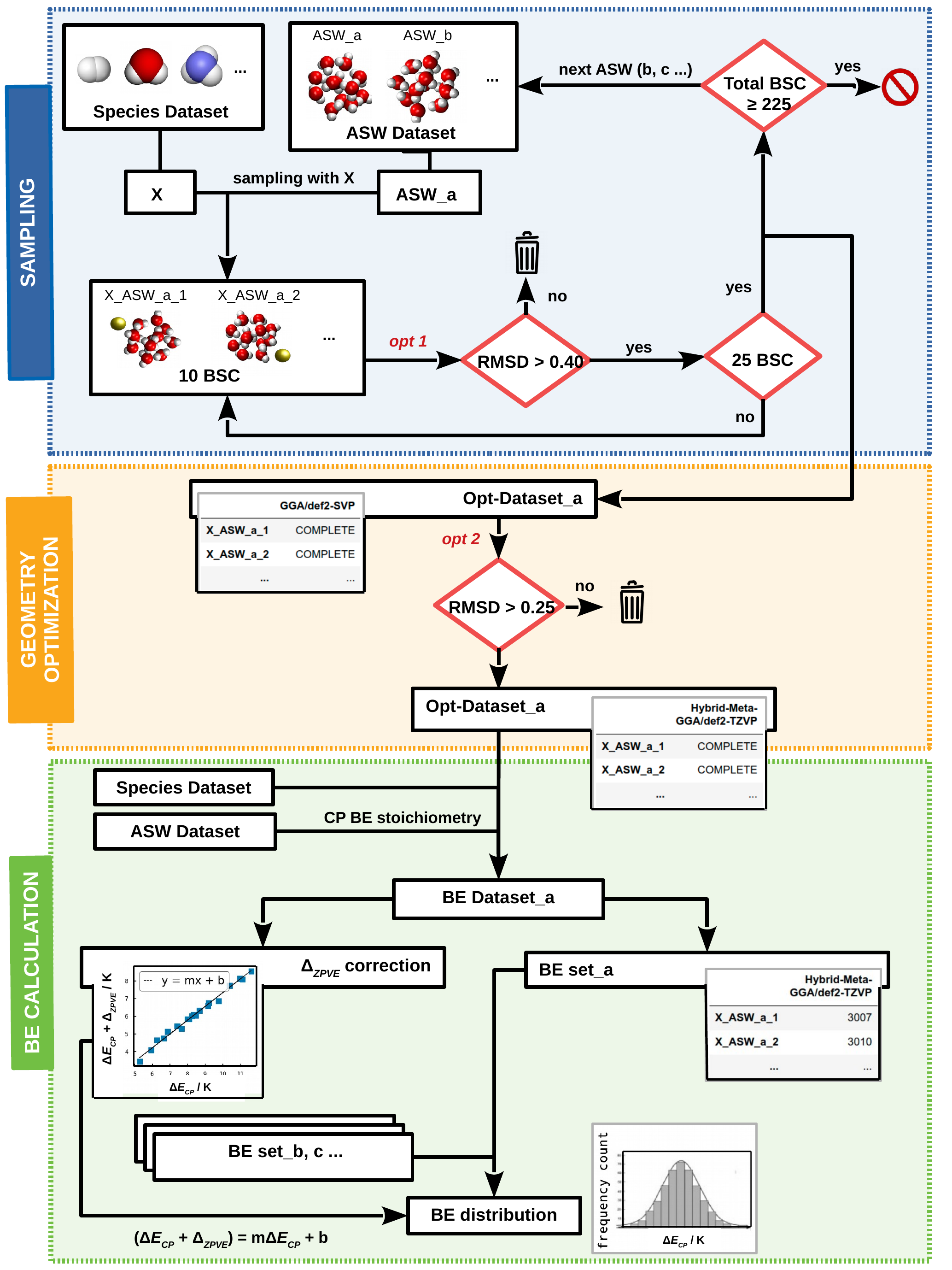}

    \caption{Three-step computational procedure used in this work for building a binding energy
    distribution. BSC stands for binding site candidate; \textit{opt 1} stands for optimization 
    at gradient generalized approximation (GGA) exchange-correlation DFT functional and 
    \textit{opt 2} for optimization at a higher level of theory that further refines the 
    geometry. The color scheme for the atoms is red for O, white for H, blue for N and 
    yellow for the generic target atom X.}
    \label{fig:pipeline}
\end{figure*}{}

 \noindent In this section we will first present each step of the computational procedure (\ref{sec:pipeline}) and then we will discuss the database results (\ref{sec:be_results}).

\subsection{Computational Procedure}\label{sec:pipeline}
\noindent The procedure we developed allows to produce ZPVE  corrected BE distributions for closed-shell and open-shell molecules. As shown in Figure \ref{fig:pipeline}, it is composed of three main steps: (1) sampling procedure, (2) geometry optimization and (3) BE calculation. 
In order to go through the procedure, we recall the reader the QCArchive data structures we introduced in \ref{sec:qcarchive}.
\subsubsection{Sampling procedure}
\noindent In order to perform the sampling procedure (Figure \ref{fig:pipeline}, blue panel labelled ``sampling")  within of the QCArchive environment, 
both the ASW clusters and the target molecules have to be stored in collection objects 
(ASW Dataset and Species Dataset). The initial molecular geometries contained in the Species Dataset are drawn 
from the Pubchem library,  which can be accessed directly from the QCArchive environment.
The sampling procedure is carried out at BLYP/def2-SVP level of theory, and consists of extracting one 
ASW structure at a time from the ASW Dataset and sample it with the target molecule X.
The sampling algorithm places the center of mass of both species on the origin of the system coordinates,  
and displaces the species X around the surface randomly within a range of distances which maximizes 
the chance of finding a binding site on the surface (within 2.5 \text{\AA} from the surface).
Starting with ice cluster \ce{ASW_a}, several groups of 
10  \ce{ASW_a-X} binding site candidates (BSC) are generated. These are optimized (\textit{opt1}) and 
filtered according to geometrical criteria, such that only the structures of RMSD $\geq$ 0.40~\AA \, with respect to previously found BSC are stored,  until 25 BSC is reached or no more new BSC are found. 
This procedure is repeated on a second cluster \ce{ASW_b}  until reaching  a total of at 
least 225 \ce{ASW-X} equilibrium structures, distributed among 12-15 ASW clusters.

\subsubsection{Geometry Optimization}
\noindent In this step (Figure \ref{fig:pipeline}, yellow panel labelled ``geometry optimization"), the BSCs
previously obtained, 
are further optimized at a more accurate level of theory,  
such as a hybrid or meta-hybrid functional with a triple-$\zeta$ basis set. According to geometry benchmark results, reported in Appendix (\ref{sec:bench_results}), 
using a computationally affordable HF-3c/MINIX \citep{sure2013}  model chemistry can also be a good
option for obtaining a refined equilibrium geometry.
In the next step, BEs will be calculated carrying out single point energy computations on the BSCs, employing a DFT functional selected through an extensive energy benchmark against a CCSD(T)/CBS reference (see Appendix \ref{sec:bench_results}).   

\subsubsection{Binding energy calculation}
\noindent The final part of the procedure (Figure \ref{fig:pipeline}, green panel labelled ``BE calculation") is the computation of BE values 
and the assembly of a ZPVE corrected BE distribution. To do so, first, the 
optimized structures are filtered with geometry criteria (RMSD $\geq$ 0.25 \AA) 
to make sure that  all binding sites on the ASW cluster are unique. The resulting
equilibrium structures are included into a BE \textit{Dataset} collection, together 
with the optimized target molecule and water cluster to create 
the stoichiometry of a BE including the counterpoise correction for the BSSE error 
(see eq. \ref{eq:bsse_1}, \ref{eq:bsse_2}). Once a BE \textit{Dataset} for \ce{ASW_a-X} is generated, it
contains all the fragments necessary to compute the BSSE corrected BE values  
on \ce{ASW_a} (BE $\mathrm{set\_a}$). Analogously, we collect a set of BEs for each 
of the sampled clusters. Assuming the clusters share common morphological characteristics, 
as they originate from a single \textit{ab initio} molecular dynamics trajectory and are annealed in the same way, the 
BEs collected  are considered as a single BE distribution of
the target molecule on the ice mantle model. We then correct the values by adding 
$\Delta_{ZPVE}$ to the BE. Due to computational cost, we compute the Hessian for the 
elements of a single BE Dataset (e.g.,  \ce{ASW_a-X} corresponding to one sampled water 
cluster in our set), and use a linear model to correlate $\Delta E_{CP}$ and $\Delta E_{CP} + 
\Delta_{ZPVE}$ (see Appendix, \ref{sec:ZPVE}). Finally, the correction factors are applied 
to all the computed BEs to obtain a ZPVE corrected BE distribution. The source code 
of the BEEP protocol and scripts to generate the data can be found in 
\url{www.github.com/QCMM/beep}.

\subsection{Binding energy distributions}\label{sec:be_results}

\begin{figure*}[ht]
    \centering
    \includegraphics[angle=0, width =0.93\linewidth]{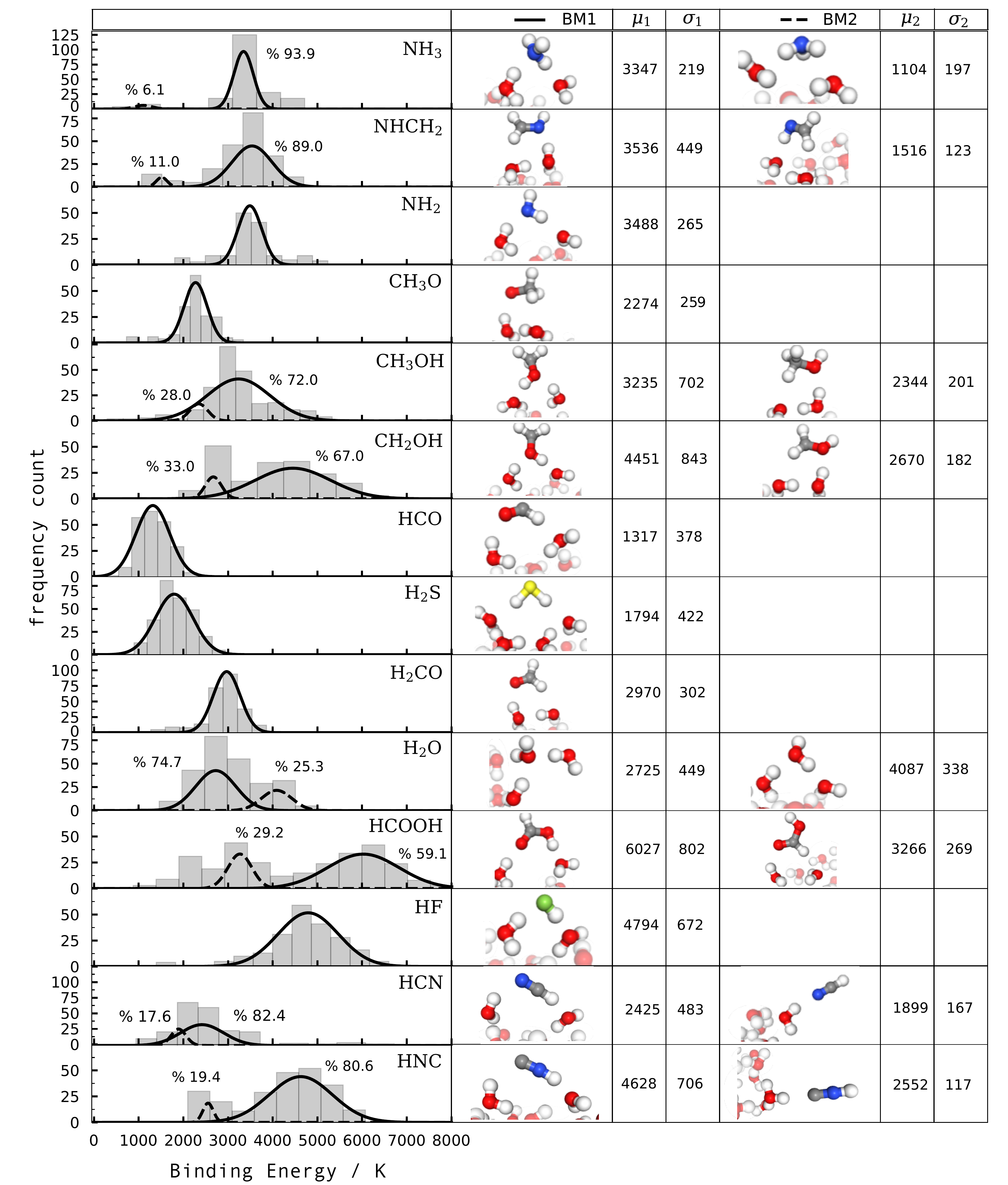}

    \caption{Binding energy distributions for Group H, \ce{ASW-X} systems, using HF-3c/MINIX geometries and including ZPVE correction. According to the benchmark results, the energy has been computed at $\omega-$PBE/def2-TZVP level of theory for all species except \ce{HNC} (B97-2/def2-TZVP), \ce{H2CO} (CAM-B3LYP/def2-TZVP),  \ce{CH3OH} (TPSSH/def2-TZVP), \ce{HF} and \ce{HCN} (MPWB1K/def2-TZVP). D3BJ dispersion correction has been applied to all DFT energies. Each identified binding mode has been fitted with a Gaussian function, using a bootstrap method (see Appendix, \ref{sec:boot}). Mean ($\mu$) and standard deviation ($\sigma$) of the
Gaussian fit are reported for the main binding mode (BM1, solid line) and the minor binding mode (BM2, dashed line). The numbers on the plot represent the percentage of minimum energy structures that belong to a specific mode. Columns 2 and 5 report a graphic representation of an example of BM1 and BM2. 
    The atoms in proximity of the binding site have been highlighted. 
    The color scheme for the atoms is red for O, grey for C, white for H, blue for N, yellow for S and green for F.}
    \label{fig: Group_H}
\end{figure*}{}

 \begin{figure*}[ht]
    \centering
     \includegraphics[angle=0, width = 0.9\linewidth]{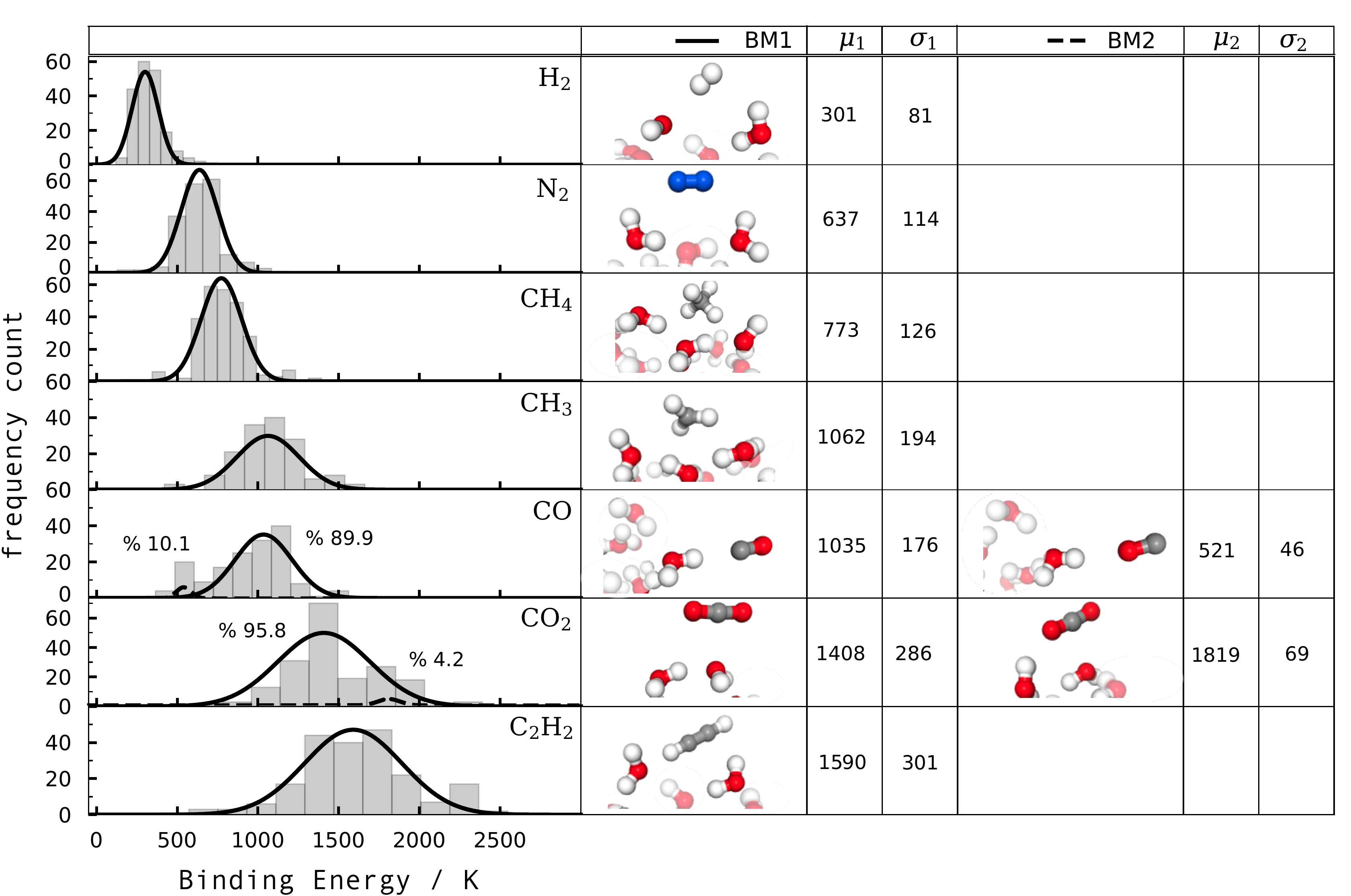}
    \caption{Binding energy distributions for Group D, \ce{ASW-X} systems, using HF-3c/MINIX geometries except for CO (M05/def2-TZVP). ZPVE correction has been included only for \ce{C2H2}, see \ref{sec:be_results}. The energy has been computed at $\omega$-PBE/def2-TZVP level of theory for all species except \ce{CH4} (TPSSH/def2-TZVP). D3BJ dispersion correction has been applied to all DFT energies. See Figure \ref{fig: Group_H} caption for further details.}
    \label{fig:Group_D}
\end{figure*}{}
\noindent We divided the molecular species in two groups according to the nature of the interaction with the ice surface. Group D accounts for interactions dominated by dispersion, while molecules in Group H predominantly bind through 
hydrogen bonds.
We computed 21 binding energy distributions for closed-shell and open-shell molecules, reported in Figure \ref{fig: Group_H} for Group H and Figure \ref{fig:Group_D} for Group D.
The equilibrium geometry is
of HF-3c/MINIX quality, 
as we probed it to be a cost-effective alternative to the more expensive DFT methods (see Appendix, \ref{sec:bench_results}).
For CO species, the geometry is M05/def2-TZVP \citep{zhao2005}, as 
HF-3c failed to properly describe the binding sites.
We computed the ZPVE correction at the HF-3c/MINIX level of theory  for Group H, 
while for most of the molecules in Group D we could not apply the linear model 
we used to derive the correction factors, due to poor correlation. This could be 
attributed to the inadequacy of the harmonic approximation to correctly describe 
the potential energy well. Notwithstanding, the correction value for Group D molecule 
is small enough to fall within the accuracy of the method. 
The BE values were computed using the best performing DFT functional from the 
energy benchmark for each molecule (see Appendix, \ref{sec:bench_results}). If no benchmark value 
is present, we used the best performing functional for each group. 

\noindent Finally, while multibinding energies approaches have been recently proposed \citep{grassi2020a}, we decided to also provide a single BE value, representative of the entire distribution, to accommodate the usage of our calculations in standard chemical models.
For this purpose, we obtained the mean BE ($\mu$) and standard deviation ($\sigma$) by fitting a Gaussian function to the distribution 
using a bootstrap method (Appendix \ref{sec:boot}). We carried out binding mode analyses
in order to identify different binding motifs which are labelled in the 
figure, along with their percentage and their $\mu$ and $\sigma$ values. An example of binding mode analysis and details about selected bond parameters can be found in a \href{https://github.com/QCMM/beep/tree/main/tutorial/data_query}{Jupyter Notebook in the BEEP GitHub repository}.

\subsubsection{Group H: Hydrogen bonded structures}\label{sec:group_H}
\noindent Figure \ref{fig: Group_H} shows BE distributions of molecules in Group H. 
These molecules are mostly bound  through electrostatic interactions in the form of 
hydrogen bonds and therefore present a strong interaction with the ASW surface. 
This is reflected in  the BE values which are in the range of 1000 to 8000 K. 
It is worth noting that several species exhibit two distinct distributions. 
For \ce{NH3} and \ce{NHCH2}, there is a main binding mode ($\mu_1$ $\sim$ 3400 K) 
where the molecule bridge two water molecules via a double hydrogen bond.
In the minor binding mode
($\mu_2$ $\sim$ 1400 K), 
the surface water molecules act solely as hydrogen-bond acceptors, resulting in a lower mean BE. 
For \ce{CH3OH} and its radical species (\ce{CH2OH}), the main 
binding mode is the surface interaction via the 
OH moiety ($\mu_1$ = 3235 and 4451 K, respectively); while in the minor
binding mode the methyl end of the molecule also participates in the interaction. In the distribution 
of the \ce{CH3O} radical, we found a single binding mode, corresponding to the 
less energetically favorable interaction  where both the oxygen and the methyl take part.
This is consistent with the inability of this radical to form a donor-type hydrogen bond.
Due to its lack of symmetry, the formic acid presents a  
rather complex BE distribution with two different components, spanning a range of 
almost 7000 K. The minor mode present dangling OH-bonds as in the case of 
the methanol species. 
Regarding the water molecule, a closer inspection of the 
binding modes shows a varied scenario where the molecule 
establishes a single ($\mu_1$ = 2725 K) or 
double ($\mu_2$ = 4087 K) hydrogen bond to the surface.
Even though the former occurs more often during the sampling 
procedure, the majority of the water molecules that compose the ASW surface form two hydrogen bonds, therefore water surface evaporation would mostly fall within the higher BE regime. 
The halogen (HF) and pseudo-halogen (HNC, HCN) molecules have a high standard 
deviation ($\sigma$ $\sim$ 600 K) which  reflects a high capacity of insertion into the ASW
environment. This is especially seen in the HF case, in which the molecule is 
easily inserted into the hydrogen bond network, forming strong hydrogen bonds with 
the water surface, as we have shown in a previous work \citep{bovolenta2020a}.
Both HCN and HNC species exhibit two binding modes. In the main one ($\mu_1$ = 2425 and 4628 K, respectively), the molecules establish a double hydrogen bonding interaction with the surface.     
We also studied the HCl molecule, but it does not have a BE distribution 
as it dissociates to its ionic components in the majority of the binding sites, 
as also pointed out in the recent work of \cite{ferrero2020}.
Finally, it is worth noting
that the ZPVE correction can significantly reduce the BEs,  in some cases up 
to 25\% of the non-corrected value.

\subsubsection{Group D: structures bound by dispersion}\label{sec:group_D}
Figure \ref{fig:Group_D} shows the BE distributions of Group D.  In order to identify the molecules 
that belong to this group, we compared the BE distributions obtained 
with and without including D3BJ dispersion correction to the energy computation. For molecules in Group D, the dispersion interaction is fundamental in order to achieve an 
attractive interaction with the water surface (see Appendix, \ref{sec:dispersion}). 
They are mainly homonuclear or highly symmetric molecules. The mean BE values range between 300 and 1800 K and are significantly lower than in the Group H molecules. 
Furthermore, the standard deviation is also less than in Group H molecules, which is 
consistent with a smaller capacity of the molecule to deform the binding site
environment. 
Most molecules therefore present a single binding motif.
An outlier is  CO, since its BE distribution reveals two 
distinct binding modes: a weak interaction where the CO molecule is bound to the 
surface via an electrostatically unfavorable CO--H interaction ($\mu_2$ = 521 K) and a
second, 
which comprises 89.9 \% of the structures, and involves the
C- extremity of the molecule ($\mu_1$ = 1035 K). 
The other molecule that presents more than one binding mode is \ce{CO2}. 
In the highest BE motif the \ce{CO2} interacts with the surface through both the C and one of the O atoms of the molecule ($\mu_2$ = 1819 K).

\section{Comparison with experimental results and previous theoretical studies}\label{sec:comparison}

\begin{table*}[ht]
 \begin{threeparttable}
\small
     \centering
     \caption{Comparison with data from the literature. The first column reports the molecules, column 2 and 3 our results: the mean of the predominant binding modes identified ($\mu_1$, $\mu_2$) and the highest BE value of each distribution (Max).
Column 4 to 5 reports He et al. experimental results at different surface coverage ($\theta$); columns 6 to 8 BEs computed in theoretical studies, columns 9 and 10 the values present in the astrochemical databases KIDA and UMIST. Units are in K and the references are listed in the notes below.}
      \label{tbl:comparison}
     \begin{tabular*}{\textwidth}{@{\extracolsep{\fill}}l  c c c  c c c c c c }
\hline
&  \multicolumn{2}{c}{\textbf{BEEP} (ASW)} & \multicolumn{2}{c}{\textbf{He} (np-ASW)\tnote{a}}  &  \multirow{2}{*}{\textbf{Das}\tnote{c}} &  \multicolumn{2}{c}{\textbf{Ferrero} (ASW)\tnote{d}} &  \multirow{2}{*}{\textbf{KIDA}\tnote{e}} &  \multirow{2}{*}{\textbf{UMIST}\tnote{f}}\\  

 & \multirow{1}{*}{\textbf{$\boldsymbol\mu_1$, $\boldsymbol\mu_2$}}  &  \multirow{1}{*}{\textbf{Max}} &    \multirow{1}{*}{$\boldsymbol\theta$ $\boldsymbol\simeq$ \textbf{1ML}} & \multirow{1}{*}{$\boldsymbol\theta$ $\boldsymbol\rightarrow$ \textbf{0}} &     &  \multirow{1}{*}{\textbf{Min}} & \multirow{1}{*}{\textbf{Max}} &  &  \\  
\hline
 \ce{H2} & 310 & 660     & 322  & 505                                    & 528  & 226 &  431  & 440 & 430 \\ 
 \ce{N2} & 637& 1189     & 790  & 1320                              & 900  & 760 &  1458  & 1100    & 790 \\
\ce{CH4}   & 773& 1393   & 1100 & 1600                               & 1327 & 914 &  1674  & 960    & 1090 \\
 \ce{CH3}  & 1062 & 1662 &      &                                   & 1322 & 1109 & 1654     & 1600 & 1175 \\
\ce{CO} & 1035 & 1561    & 870 & 1600                                  & 1263 & 1109 & 1869 & 1300  & 1150\\    
\ce{CO2} & 1408, 1819 & 2389  & \multicolumn{2}{c}{2320\tnote{h}} & 2293 & 1489 & 2948   & 2600     & 2990 \\    
\ce{C2H2} & 1590 & 2547 &  &                                      & 2593 &          &     & 2587    & 2587   \\    
\ce{NH3}    & 3347, 1104 & 4715 & &                           & 3825 & 4314 & 7549       & 5500     & 5534 \\
\ce{NHCH2}    & 3536, 1516  & 4695 & &                         & 3354 &      & & 5534\tnote{m}     & 3428 \\
\ce{NH2}    & 3488 & 5235 & &                              & 3240 & 2876 & 4459          & 3200     & 3956 \\
\ce{CH3O}    & 2274 & 3343   & & &  & &                                                  & 4400     & 5080 \\
  \ce{CH3OH} & 3235, 2344 & 5331 & &                          & 4368 & 3770 & 8618       & 5000     & 4930 \\
  \ce{CH2OH} & 4451, 2670& 6594 & &                           & 4772 &             &     & 4400     & 5084 \\
  \ce{HCO} & 1317 & 3764 & &                                  & 1857 & 1315 & 3081       & 2400     & 1600 \\
  \ce{H2S} & 1794 & 2940 & &                                  & 2556 & 2291 & 3338       & 2700     & 2743 \\
  \ce{H2CO} & 2970 & 3800 & &                                     & 3242 & 3071 & 6194   & 4500     & 2050 \\
  \ce{H2O} & 2725, 4087 & 4885 & &                                  & 2670 & 3605 & 6111 & 5600     & 4800  \\
  \ce{HCOOH} & 6027, 3266 & 8044 & &                & 3483 & 5382 & 10559    & 5570\tnote{n}       & 5000 \\
   HF & 4794 & 6500 & &                                          & 5540 &              & & 7500     &  \\
  HCl & \tnote{g} & \tnote{g} & &                 & 3924 & \tnote{g} & \tnote{g} & 5172     & 900 \\
 HCN & 2425, 1899 & 4252 & &                                       & 2352 & 2496 & 6337        & 3700     & 2050 \\
 HNC & 4628, 2552 & 6570 & &                                   & 5225 &              &   & 3800     & 2050 \\
\hline
 \end{tabular*}
 \begin{tablenotes}
 \footnotesize 
\item[a] \cite{he2016a}; \item[c] \cite{das2018}; \item[d] \cite{ferrero2020}; \item[e] \cite{Wakelam2017}; \item[f] \cite{mcelroy2013}; \item[g] HCl molecules dissociate; \item[h] coverage insensitive; \item[m] \cite{ruaud2015};  \item[n] \cite{collings2004}. 
\end{tablenotes}
\end{threeparttable}
\end{table*}

\begin{figure}[h]
    \centering
    \includegraphics[angle=0, width = 0.7\linewidth]{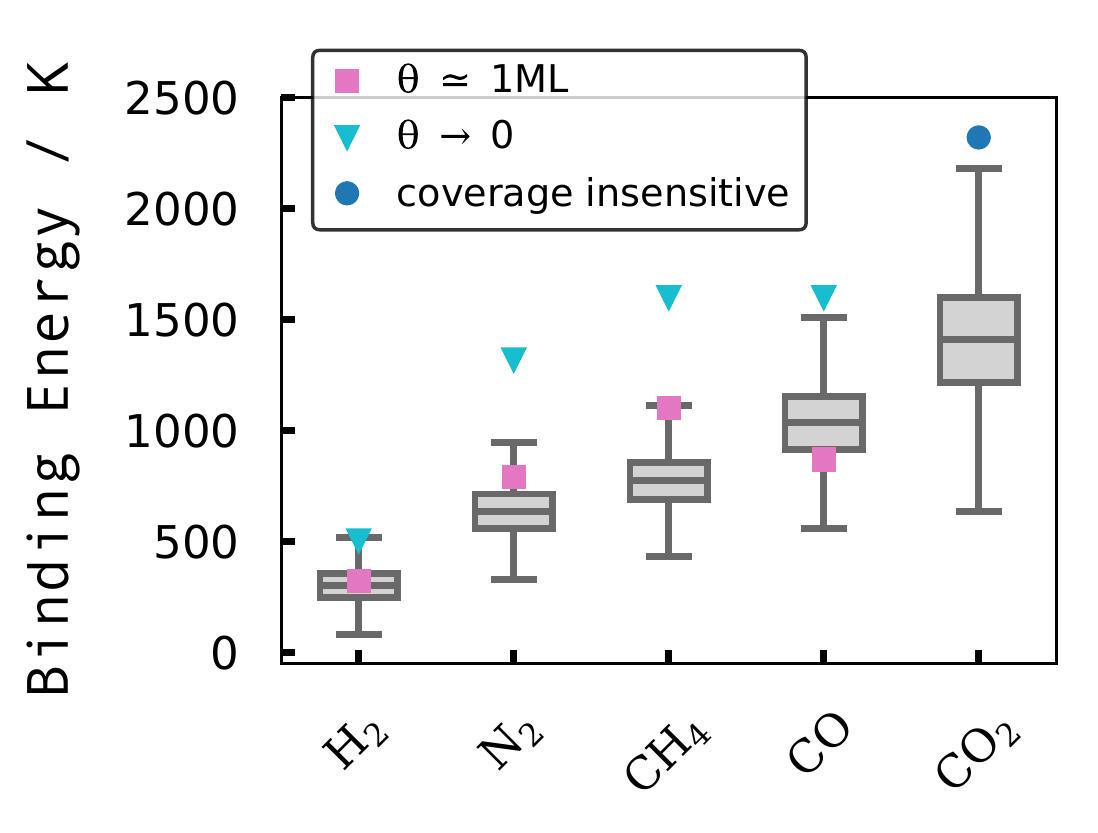}
    
    \caption{Box plot comparison between BE distributions presented in this work and  \cite{he2016a} experimental results. The box plot median corresponds to the mean of the main binding mode we identified.}
    \label{fig:he}
\end{figure}

\begin{figure*}[h]
    \centering
    \includegraphics[angle=0, width = 
 1.05\linewidth]{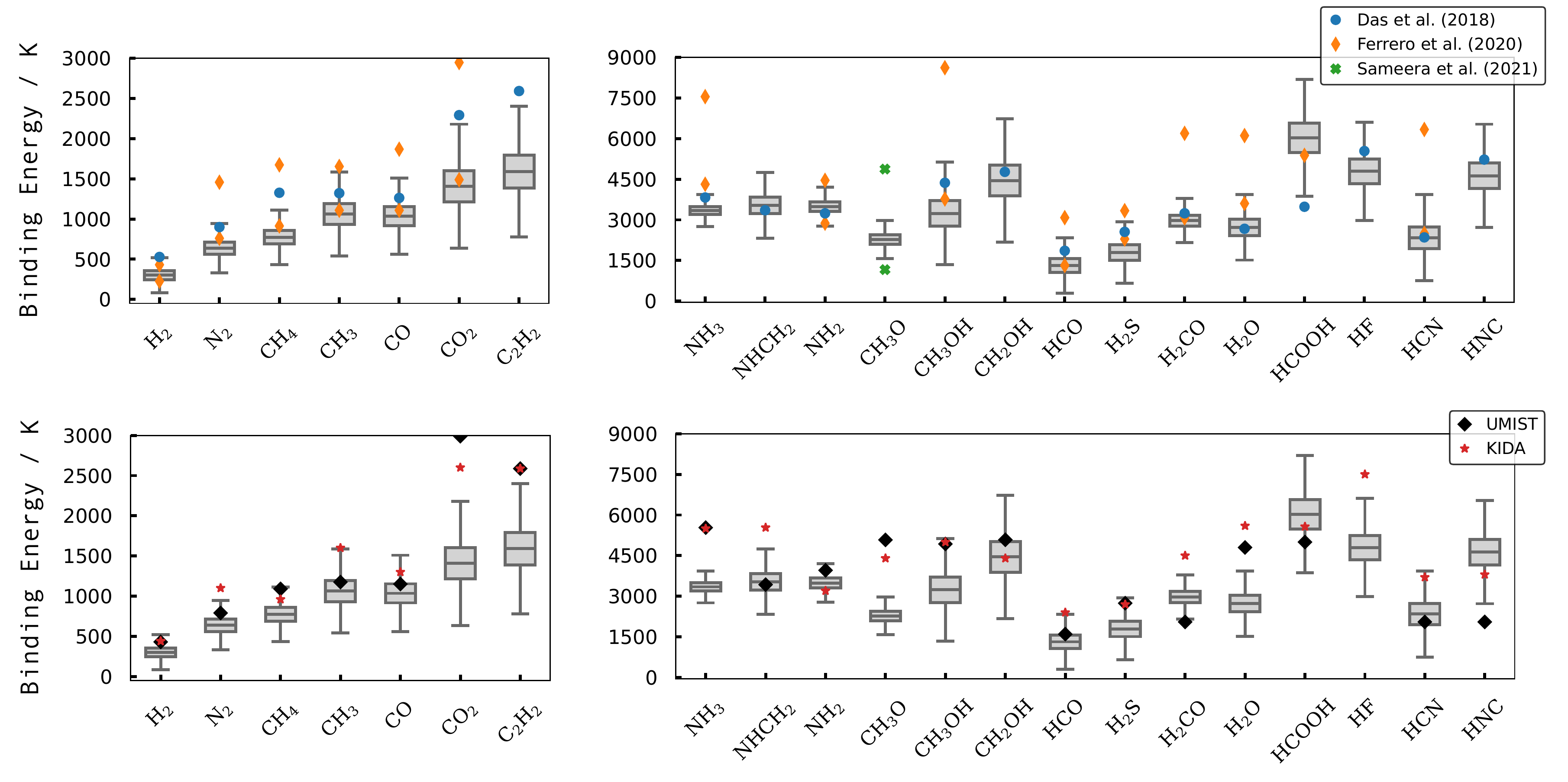}
 
  \caption{Box plot comparison between BE distributions computed in this work, considering only the main binding mode, and BEs present in previous theoretical studies (\cite{das2018,ferrero2020, sameera2021}) upper panels, and existing databases BE data: KIDA \citep{Wakelam2017}, UMIST \citep{mcelroy2013}, lower panels.}
    \label{fig:data_theo}
\end{figure*}

\noindent We compared our BE values with available experimental results, previous
theoretical studies and existing astrochemical databases.
Making a meaningful comparison of calculated BEs with experimental data is challenging, due to the variety of conditions under which the experiments are performed. In addition, the experimental data strongly depend on the pre-exponential factor used in the Polanyi-Wigner equation employed to derive the BEs \citep[see][]{minissale2022} and the fitting procedure for obtaining BEs from TPD 
temperature curves.
We decided to take into account the work of \cite{he2016a} where they
presented TPD measurement of BEs of relatively simple molecules (\ce{N2} , \ce{H2},
\ce{CO}, \ce{CH4}, and \ce{CO2}) on a non-porous ASW  (np-ASW) surface at monolayer (ML)
and submonolayer coverage. 
In He et al. experiments, it is possible to distinguish between two situations in terms of the
coverage ($\theta$) of the target molecule on the surface. The low
coverage limit ($\theta$ $\rightarrow$ 0),  
represents a situation in which mostly
the binding sites of high BE would be occupied, corresponding to 
the high energy tail of the
BE distribution.  
On the other hand, BE values obtained at the monolayer regime
($\theta$ $\simeq$ 1 ML) can be related with the mean of our BE distribution, where a 
variety of adsorption sites with different energies are occupied.
The comparison between our results and their low coverage and ML regime BEs is shown in Figure \ref{fig:he}, where our BE distributions are represented as box plots. 
Overall, the experimental results of these limiting coverage cases coincide  well
with the computational values obtained in this work.
The comparison is particularly good for \ce{H2}, \ce{N2} and \ce{CO},
(a difference of $<$ 155 K in the low coverage regime
and  $<$ 170 K in the ML regime) while the error  for
\ce{CH4} is larger (a difference in low coverage regime of 207 K, and a difference in ML regime of 337 K). 
In light of these results, we conclude that our approach of sampling a number of independent 
ASW clusters of a limited size (22 water molecules) allows to reproduce 
the statistical nature of the interaction of those molecules with an actual ice surface. \\
Regarding the comparison with previously reported theoretical values, we took into account the works of \cite{das2018} and \cite{ferrero2020} (Figure \ref{fig:data_theo}, upper panel). 
Das et al. built a database of BE values for \ce{W_4-X} systems at MP2/aug-cc-pVDZ level of theory without  
correction for BSSE  and nor for ZPVE. Also, the existence
of multiple binding sites is not considered. Regarding Group H, in most cases 
Das' values  fall within the range of energies we found for the same systems.
For molecules in Group D, Das' values mostly overestimate ours. This is consistent with the lack
of BSSE correction that has an important effect on the final BE values for this 
group (BSSE correction $\sim$ 100-250 K in our BE results).\\
Recently, Ferrero et al. proposed a new set of BE values, computed at DFT/A-VTZ* level, including ZPVE and BSSE correction. 
Their single ASW model slab contains a cavity that allowed them to explore up to 8 different 
binding sites.
The aim of their work was different than ours 
inasmuch as they tried to obtain a range of possible BE values and not a full distribution. 
   Their lower BE fall within our distribution for most of the systems, 
but their BEs are in average higher than the ones presented here. A possible reason for higher BE is the shape of the water cluster, which is essentially  a nano-cavity 
and, as recently pointed out  \citep[see][]{rimola2018,enrique-romero2019, bovolenta2020a}, 
the presence of cavities notably increases the BEs, as they offer more favourable interaction 
sites for the molecule on the surface.  It is still uncertain to which extent the real ASW 
ice-mantle surface contains such defects, and therefore how statistically relevant they are 
for our aim of obtaining a full distribution of BE. 
In a recent work, \cite{germain2022} constructed an ASW cluster containing  
200 water molecules  and computed the BE distribution for the \ce{NH3} molecule using the GFN2
semiempirical tight-binding method and a GFN-FF force field. Their cluster model contains some nano-cavities and
the reported mean BE (4089 K) falls  within 
\cite{ferrero2020} average value (5932 K)
and the one reported in this work (3347 K).
\noindent Regarding the \ce{CH3O} radical, we took into account for comparison the recent work of \cite{sameera2021}. They used 10 molecular-dynamic generated ASW
structural models composed of 162 water molecules, which have been sampled with the target \ce{CH3O}. The resulting 10 BEs have been computed using the two-layer 
ONIOM(QM:MM) approach, at $\omega$-B97XD/def2-TZVP 
\citep{chai2008} level of theory including ZPVE 
correction; we reported their minimum and maximum values 
in Figure \ref{fig:data_theo}, upper panel. They 
identified a wide range of energy (1160 - 4874 K), that 
encompasses the values of our distribution.
Finally, in Fig. \ref{fig:data_theo}, lower panel, we show the comparison of our 
data with the largely used KIDA and UMIST databases values. They mostly fall in 
the range of our BE distributions, except for some specific cases, where the 
agreement is poor 
(\ce{CO2} and \ce{NH3} among them). 
These KIDA values are mostly based on the BE calculated in \cite{Wakelam2017} 
using a semi-empirical model consisting of a linear fit between the BEs on water 
monomers and experimental values on ASW surfaces. The BEs calculated using this model 
tend to overestimate  our average values for both Group H and D. It is important 
to consider that
their model is based on results from different experimental setups, which makes a meaningful comparison difficult.

{}

{}

\begin{figure*}
    \centering
    \includegraphics[angle=0, width = 0.9\linewidth]{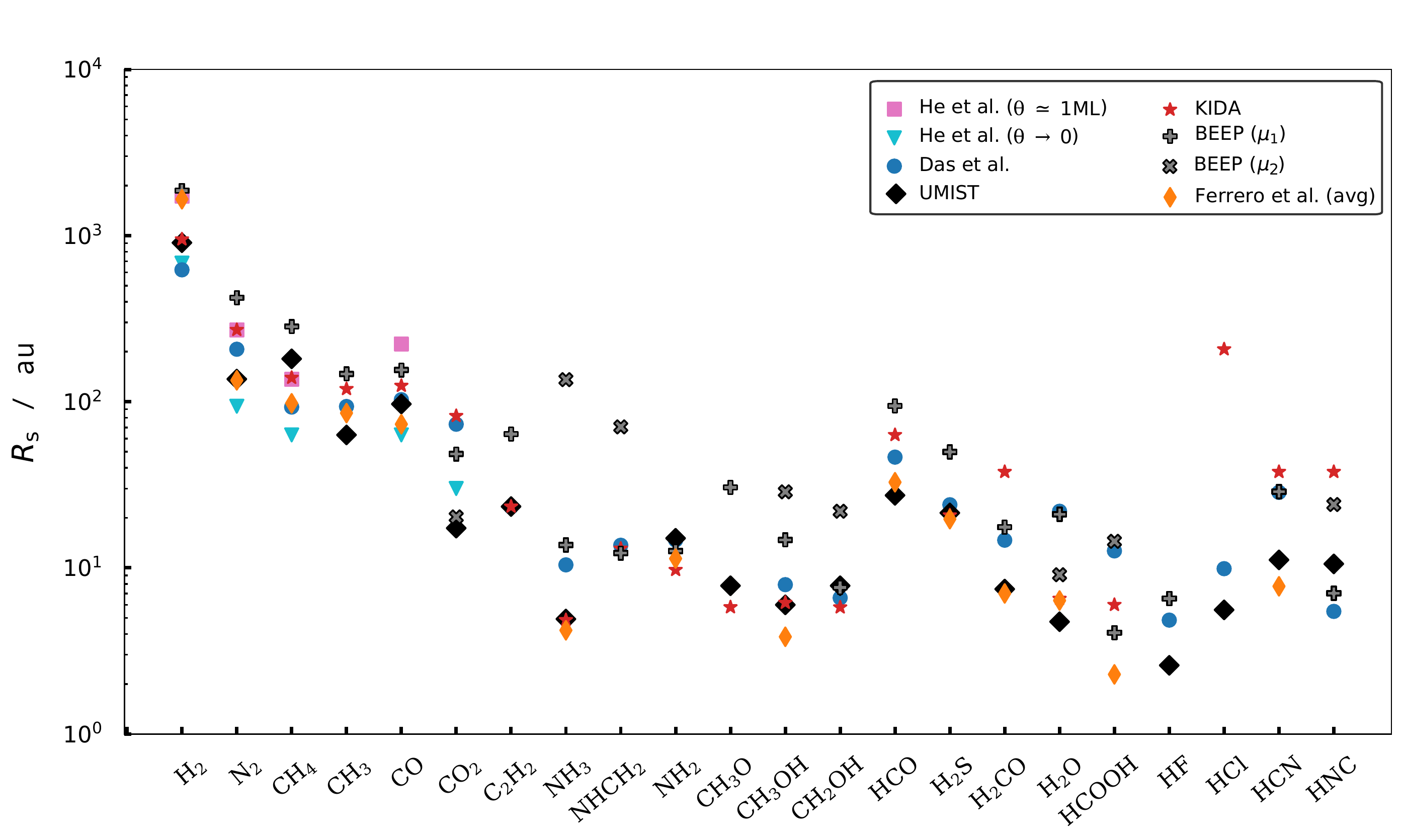}
    \caption{Sublimation radius on the midplane of a typical protoplanety disk for different species obtained by employing the binding energies obtained in this work and compared with values available in literature. For the sake of comparison, Ferrero~et~al. is the average between their upper and lower limits. Cf. the references in Fig.~\ref{fig:data_theo}.}
    \label{fig:astro_app}
\end{figure*}

\section{Astrophysical implications}
\noindent 
When comparing results between chemical experiments and quantum chemistry computations, a difference of 0.2-0.3~kcal\,mol$^{-1}$ (corresponding approximately to 100-150~K) in the final BE is not substantial. However, for astrochemistry modelling, a few tens of Kelvins could largely affect the final outcome.   
The molecular desorption is described by the Polanyi-Wigner equation, where its dependence on the exponential of the BE plays a crucial role in determining the efficiency of the process. To show this effect on a realistic, yet idealized, astrophysical case, we have calculated the sublimation radius in a protoplanetary disc (i.e., the so-called snow line) by equating the desorption and the viscous time, and finding the corresponding radius \citep[see e.g.][]{grassi2020a}. The evaporation time is defined as $t_\mathrm{des}(R)=\nu_0 \exp[{\rm \Delta E_b}/k_{\rm B}T_\mathrm{d}(R)]$, with $\nu_0=10^{12}$\,s$^{-1}$, $k_{\rm B}$, and $T_\mathrm{d}$ respectively the Debye frequency, the Boltzmann constant, and the dust temperature at a given radius $R$. The viscous time is $t_\nu(R) = R^2  \nu^{-1}(R)$, where $\nu(R) = \alpha\, c_{\rm s}^2(R) \Omega_{\rm K}^{-1}(R)$ is the viscosity, assuming an $\alpha$-viscous prescription with $\alpha=10^{-2}$, and $c_{\rm s}$ the speed of sound and $\Omega_{\rm K}$ the Keplerian angular frequency. By means of the bisection method, we solve $t_\mathrm{des}(R) = t_\nu(R)$ for $R$, that, assuming a temperature radial profile of $T(R)\propto R^{-0.5}$, corresponds to ${\varphi_1 \ln\left({\varphi_2 R}\right) - {\rm \Delta E_b} \sqrt{R} = 0}$, with $\varphi_1$ and $\varphi_2$ containing all the constant terms (see Appendix \ref{sec:framework_astro} for more details). The results are reported in Fig.\,\ref{fig:astro_app}, where the root of the aforementioned transcendental equation is defined sublimation radius, $R_{\rm s}$. As expected, the position of the snow lines is affected by the assumed BE up to approximately an order of magnitude in the worst cases. For water, one of the most important molecule involved in the process of planet formation, we obtain $R = 6$\,au for the BE computed by Ferrero et al. (the average between their reported maximum and minimum BE values),   
and 9\,au for the ice evaporation binding mode mean value ($\mu_2$) computed in this work. A similar effect is reported for CO, with up to a factor of three in the final radius. 
Other species, like molecular hydrogen, show larger differences; however we do not expect them to form observable snow lines, since they are involved in other chemical processes that are not captured by our simplified disk model, but we report them anyway for the sake of completeness. With a binding site distribution, within the framework of this idealized disk model, we expect to observe a smoothed snow line, determined by the interplay between the temperature density profile, and the BE distribution. Increasing the distance from the star, and consequently decreasing the dust temperature, the number of available lower-energy binding sites will grow, depending on the broadening of the distribution. Conversely, a single BE will produce a sharp transition. An accurate determination of the BE is then fundamental to quantitatively assess quantities like snow line positions in planet-forming regions and evaporation fronts during star-formation.

\section{Database features, accessibility and use-cases}\label{sec:features}
\noindent Due to the nature of QCArchive Databases, BEEP is extendable to an increasingly 
large number of molecules. Moreover, different cluster surface models
of different sizes and composition (e.g., different ice mixtures) can be  easily added to the platform  environment 
and used to produce new BE distribution data. At the moment, the BEEP platform can be accessed with a username and password, 
which are provided in the Appendix \ref{sec:BEEP_access}. 
This allows the user to query 
the database for BE, binding site structures and many other properties. 
To make the access to the database a user-friendly experience, we included a 
Python module that allows to query the data without having to know the QCArchive 
syntax.  The core of this  Python module is the \textit{BindingParadise} class 
that is initialized with the user's  credentials and allows to set molecules and obtain all the related  BE data.  In the GitHub  repository (\url{www.github.com/qcmm/beep}) we 
included an example jupyter-notebook to showcase the different query options.
The libraries to compute and store a BE distribution are also contained in 
the Python module. In principle, any researcher can install the module to run the 
software and  spin up a QCFractal server to store its own BE data. However,
our idea is to make this a collaborative endeavour in which different researchers 
use the proposed protocol to generate new BE data and store it in our open BEEP 
database. This allows us to expand the database in terms of new ice models and a more extensive BE catalog with more computed molecules. 
The database will be able to produce input files in the standard astrochemical software format, both in a single BE fashion and in more complex multibinding approaches.
A database of reproducible and accurate BEs is also a fundamental starting point to chemical reactivity studies 
and diffusion of molecules on the surface of interstellar ices, as having a potential energy map of neighbouring binding sites will be paramount in finding diffusive transition states and computing diffusion energy barriers. 

\section{Conclusion}

\noindent In this work, we present a Binding Energy Evaluation Platform (BEEP) that 
implements a protocol to compute binding energies on ASW cluster models. It 
also contains a database that allows to query the results produced by the 
protocol. BEEP consists of three highly automated steps: target molecule sampling procedure, geometry optimization of the binding site and 
binding energy computation, by means of DFT methods.  The  binding energy distributions were obtained by sampling  the
ASW model spanned by a set of 12-15 amorphized water clusters containing
22 molecules each.  
We categorize the molecules into two groups, based on their type of interaction with the surface: molecules that are bound  
primarily through hydrogen-bonds (Group H) and molecules for which dispersion interactions 
enable binding to the surface (Group D). 
We computed 21 binding energy distributions of astrophysically relevant molecules. 
Each distribution contains between 220-230 binding sites. We report mean values and standard deviation for all distributions, obtained using a Gaussian fit. Most molecules in Group H present two distributions, corresponding to different 
binding modes, 
while Group D molecules mainly have one.  
Group H molecules present stronger interactions with the surface, which is consistent with higher mean binding energies with respect to dispersion bound Group D molecules.  

We studied the effects of our new calculations on the position of the snowline of an idealized protoplanetary disk, finding that our approach might play a relevant role in determining the correct position of the sublimation front. In particular, the sublimation radius changes by a factor of a few, for example in the case of methane, up to an order of magnitude in the case of e.g.~methanol or ammonia. This suggests that accurate binding energies might have a marked effect on some of the key astrophysical observables.

BEEP is build on an
open-source platform and hence any researcher can use it to compute binding energies with a cluster based 
ice surface model. Finally, we plan to transform BEEP into a widely-used tool for standardized \textit{ab initio}  
binding energy data for astrochemical modeling and ice-grain surface processes studies.

\section*{Acknowledgments}
\begin{acknowledgments}
\noindent The computations were performed with resources provided by the Kultrun Astronomy Hybrid
Cluster hosted at the Astronomy Department, Universidad de Concepci\'on. We would like to thank Benjamin Pritchard for his guidance on the QCFractal platform. GMB gratefully acknowledges support from ANID Beca de Doctorado Nacional 21200180 and Proyecto UCO 1866 - Beneficios Movilidad 2021. SB gratefully acknowledges support by the ANID BASAL projects ACE210002 and FB210003. 
SVG
is financially supported by ANID grant 11170949.
\end{acknowledgments}

\clearpage
\bibliography{Astrochem,QCMM,PhD_Project,Astrochem6}

\appendix

\section{BEEP database access}\label{sec:BEEP_access}

\noindent The BE and binding site data generated using BEEP can be accessed using 
the Python \textit{BindingParadise} class. Refer to the GitHub repositories for
installation instructions. To initialize a class object and access the data you can use
the following credentials: 
\begin{center}
\noindent username : guest \hspace{2cm}password: pOg\_41tzuDxkTtAfjPuUq8WK5ssbnmN8QfjsApGXVYk\\
\end{center}

\noindent Examples of how to use the class with jupyter-notebook can be found in the tutorial section of our
GitHub repository at \url{www.github.com/QCMM/beep}. 

\section{Estimation of computation time}\label{sec:timings}
\noindent We report a detailed estimation of the computation time in order to produce the full BE distribution of an example molecule (constituted by around 225 binding sites), along with the  computational resources used in this work.
\begin{itemize}
\item[$\square$] Sampling procedure, carried out at BLYP/def2-SVP level of theory: 1 week, using 4 Tesla GPUs.
\item[$\square$] Geometry optimization, carried out at hybrid or meta-hybrid/def2-TZVP level of theory: 3 weeks, using 128 standard high-performance Intel Xeon CPUs. 
\item[$\square$] Geometry optimization, carried out at HF-3c/MINIX level of theory: 1 day, using 128 standard  high-performance Intel Xeon CPUs.
\item[$\square$] BE computation, carried out at hybrid or meta-hybrid/def2-TZVP level of theory: 2 days, using 40 standard high-performance Intel  Xeon CPUs.   
\end{itemize}

\section{BSSE corrected BE calculation stoichoimetry}\label{sec:BE_stoich}
\noindent In the following, we define the electronic energy of a molecule M 
in the geometry G computed with the 
basis $\gamma$ as $E_{M}^{G}(\gamma)$. Considering this notation, the BE of a molecule X with a 
basis set $\chi$ on a water cluster W with a basis set $\omega$ can be calculated as:

\begin{equation}
\Delta E_e = E_{XW}^{XW}(\chi\cup\omega) - (E_{X}^{X}(\chi) + E_{W}^{W}(\omega))
\label{eq:bsse_1}
\end{equation} 

\noindent However, when using this expression, one does not consider that the basis function centered 
at W assists in lowering the energy of fragment X and viceversa, resulting in a lower electronic energy of the supermolecule
($E_{XW}^{XW}(\chi\cup\omega)$) and hence an overestimation of the BE. This effect
is commonly known as basis set superposition error (BSSE). A way to correct for this error is the 
so-called counterpoise method (CP) \citep{boys1970}, that considers the energy of the fragments in the geometry
of the supermolecule with the basis of the respective partner. Thus the correction is 
calculated as:

\begin{equation}
\begin{split}
\Delta_{CP}  = E_{X}^{XW}(\chi\cup\omega) - E_{X}^{XW}(\chi) \\
+ E_{W}^{XW}(\chi\cup\omega) - E_{W}^{XW}(\omega)
\end{split}
\label{eq:bsse_2}
\end{equation}

\noindent Such that the resulting BE is:

\begin{equation}
\begin{split}
\Delta E_{CP}  =\Delta E_e - \Delta_{CP}
\end{split}
\label{eq:bsse}
\end{equation}

\noindent It is important to notice that at the CBS limit, the correction term
is zero since,  $\chi, \omega$ and $\chi\cup\omega$ are 
the same. 

\section{Geometry Optimization and $\Delta_{ZPVE}$ correction}\label{sec:ZPVE}
\noindent The optimization algorithm for all equilibrium structures presented in this work is 
geomeTRIC \citep{wang2016}, which uses a coordinate system especially suitable for
optimizations of non-covalently bound systems.
Due to computational cost, we computed the Hessian matrix for the binding sites of a single ASW cluster, at the level of theory of the optimization (HF-3c/MINIX), in order to obtain the Zero-Point Vibrational Energy contribution ($\Delta_{ZPVE}$) to the BE:

\begin{equation}
    \Delta_{ZPVE} = ZPVE_{XW} - (ZPVE_{X} + ZPVE_{W})
\end{equation}
with X being the target molecule, W the water cluster and XW the supermolecule.\\ 
\noindent The linear model we used to correct $\Delta E_{CP}$ is an equation in the form: 

\begin{equation}\label{eq:linear_fit2}
\Delta E_{CP}+\Delta_{ZPVE} = m \Delta E_{CP} + b, 
\end{equation}
\noindent with $m$ and $b$ being the ZPVE correction factors. A list of correction factors for each species is reported in Table \ref{tbl:zpve}. 
Finally, the factors are applied 
to the set of computed BEs for each species in order to derive the ZPVE corrected BE distribution.
An example plot with the linear model applied to \ce{H_2CO} molecule and the code we used in order to process the computed Hessian data 
can be found at \url{www.github.com/QCMM/beep}. 

\begin{table*}[ht]
 \begin{threeparttable}
\small
     \centering
     \caption{Column 1: species; column 2: average BE values calculated in this work with ZPVE correction. Columns 3-5: $\Delta_{ZPVE}$ computed at HF-3c/MINIX and correction factors (\textit{m} and \textit{b}) obtained using a linear model. All the energies are in Kelvin.}  
      \label{tbl:zpve}
     \begin{tabular*}{\textwidth}{@{\extracolsep{\fill}}l  c c c c}
     \hline

 & \multirow{2}{*}{\textbf{$\boldsymbol\mu_1$, $\boldsymbol\mu_2$}}  &  \multirow{2}{*}{\textbf{$\boldsymbol\Delta_{\textbf{\textit{ZPVE}}}$}} &    \multirow{2}{*}{\textbf{\textit{m}}} & \multirow{2}{*}{\textbf{\textit{b}}} \\  
  & & & & \\
\hline
\ce{C2H2} & 1590 & -389  & 0.803 & 0.000               \\    
\ce{NH3}    & 3347, 1104 & -951 &  0.762 & 0.142     \\
\ce{NHCH2}    & 3536, 1516   & -491 & 0.844 & 0.277 \\
\ce{CH3O}    & 2274 & -277 &  0.814 & 0.394  \\
  \ce{CH3OH} & 3235, 2344 &  -613 &   0.819 & 0.170 \\
  \ce{HCO} & 1317 & -297 & 0.723 & 0.299       \\
  \ce{H2S} & 1794 & -432 & 0.806 &  0.000          \\
  \ce{H2CO} & 2970 & -650 &   0.758 &  -0.446         \\
  \ce{H2O} & 2725, 4087 & -465 & 0.781 & 0.466       \\
  \ce{HCOOH} & 6027, 3266 & -960 & 0.899 &  -0.508   \\
   HF & 4794 & -1211 & 0.798 & 0.000     \\           
 HCN & 2425, 1899 & -494 & 0.826 & 0.000                 \\
 HNC & 4628, 2552 & -355 & 0.929 & 0.000           \\
\hline
 \end{tabular*}
 \begin{tablenotes}
 \footnotesize 

\end{tablenotes}
\end{threeparttable}
\end{table*}

\clearpage

\section{Geometry and energy benchmarks}\label{sec:bench_results}
\begin{figure*}[h]
    \centering
    \includegraphics[angle=0, width = 1\linewidth]{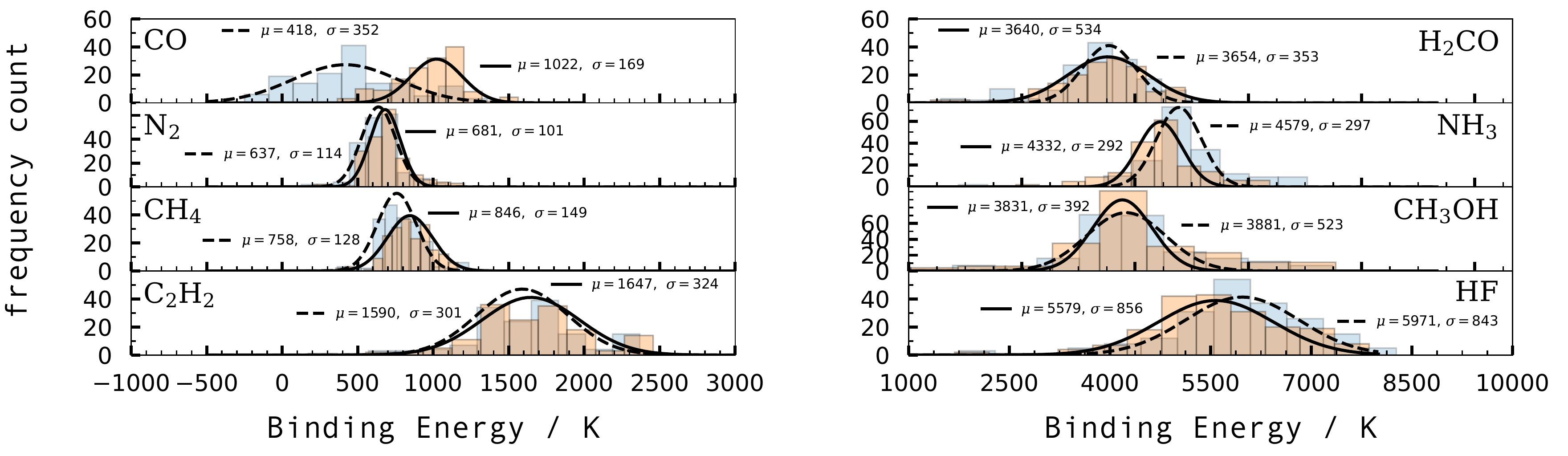}
    
    \caption{Comparison between binding energy distributions obtained using meta-hybrid GGA geometries (orange histogram, Gaussian function represented with solid line) and HF-3c geometries (blue, dashed line). Left panel: Group D; right panel: Group H. BE values are shown without ZPVE correction.}
    \label{fig:geometry_dependence}
\end{figure*}

\noindent In order to obtain the best possible equilibrium geometry at a reasonable 
computational cost, we performed a geometry benchmark on the \ce{W_{2-3}-X} systems, with X being the target molecule and W the water cluster. The benchmark has been conducted for 13 selected molecules. 
A DF-CCSD(T)-F12/cc-pVDZ-F12  geometry was used as a reference, and we 
probed 24 gradient generalized approximation (GGA)  exchange-correlation density functionals
including functionals with exact exchange (hybrid functionals), the Laplacian of the electron density (meta functionals) and long-range correction, 
paired with a def2-TZVP basis.  We also conducted an energy benchmark, using the \ce{W_{4}-X} system to compare BSSE corrected DFT BE values 
to a CCSD(T)/CBS reference energy.
The \textsc{Molpro} \citep{werner_molpro_2012} program was used for reference geometries and \textsc{Psi4} \citep{smith2020} software package was used for all energy computations. 
Table \ref{tab:bench} reports geometry benchmark results.
Generally, the meta-hybrid-GGA methods have a very good 
performance across the groups. The most dependable functionals are B3LYP\citep{becke_density_1993,lee1988} for Group D and PWB6K\citep{zhao2005a} 
for Group H, as both show an average RMSD value that is below 0.1 \AA \, with respect to the reference geometry. 
We also probed the parametrized HF-3c/MINIX \citep{sure2013} and PBEh-3c/def2-mSVP\citep{grimme2015} levels of theory.
The results are reported in the third column of Table \ref{tab:bench} and show  an average RMSD that is below 0.2 \AA\, for both groups, which is in line with the RMSD values of hybrid and 
meta-hybrid functionals.
This makes it a cost-effective alternative to the computationally 
more expensive DFT methods.\\ 
Furthermore, we evaluated the dependence of the  BE distributions on the quality of the underlying binding site geometries, comparing the BE distribution of the equilibrium structures obtained with HF-3c/MINIX to the best performing  DFT method.  
Figure \ref{fig:geometry_dependence} reports the comparison for Group D, left panel and Group H, right panel. 
For all species, the mean BE ($\mu$) presents a shift passing from DFT to parameterized methods, while the standard deviation 
($\sigma$) is mostly unchanged. The shift in the position of $\mu$ is below 400 K for all the species except CO ($\Delta\mu$ of 604 K), for which HF-3c largely underestimates the BE.
In light of these results, we conclude that the HF-3c/MINIX 
model chemistry can be used in lieu of a more expensive DFT
method, as it shows only small difference in the position and 
width of the Gaussian fit of the underlying BE distributions.\\

\begin{table}[h]
\small
    \centering
       \caption{Summary of the results of the geometry and energy benchmarks for \ce{W_{2-3}-X} (\ce{W_{2}-X} for radicals) and \ce{W_{4}-X} (\ce{W_{3}-X} for radicals) systems, respectively. The first column reports the molecules. 
       Columns 2-3
       report the performance of the best DFT functional for each group, and of HF-3c. 
       Only structures that converged (n) to the reference minima (N) were considered for the benchmark. The forth column reports reference energies calculated at CCSD(T)/CBS level of theory. The fifth column reports the Mean Absolute Error (MAE) of the best DFT functional for each group. 
       All DFT geometries and energies were computed using a def2-TZVP basis set and including D3BJ dispersion correction. 
       HF-3c method is coupled with MINIX basis set.
}
    \label{tab:bench}
\begin{tabular*}{0.78\textwidth}{@{\extracolsep{\fill}}lll lll}
\clineB{1-5}{2.5}
  & \multicolumn{2}{c}{\multirow{1}{*}{\textbf{RMSD / \AA}}}  & \multirow{1}{*}{\textbf{BEs / K}}  & \multirow{1}{*}{\textbf{MAE / K}}  \\
\cmidrule{2-5}
 \multirow{2}{*}{\textbf{Group D}} & \multirow{2}{*}{\textbf{B3LYP(n/N)}}  &  \multirow{2}{*}{\textbf{HF-3c(n/N)}}  & \multirow{2}{*}{\textbf{CCSD(T)/CBS}}  & \multirow{2}{*}{\textbf{$\boldsymbol\omega$-PBE}}  \\  
  & &  \\
 \cmidrule{1-5} 
 \ce{H2} &  0.11 (3/6) & 0.14 (5/6) &   320, 	116 &   19 \\
 \ce{CO}   &  0.12 (7/7) & 0.22 (5/7)     & 950, 870, 791 & 10 \\
    \ce{CH4} &   0.07 (2/2) & 0.14 (2/2)    &  712 &  74  \\
 \ce{CH3}  &  0.09 (1/2) & 0.12 (1/2)  &    821, 	824 & 45 \\
 \ce{N2}     &        0.13 (3/3) & 0.26 (3/3) &  &  \\
 \hline
  Average &  0.10 (16/20) & 0.18 (16/20) & & 37   \\
 \hline
 \multirow{2}{*}{\textbf{Group H}}      & \multirow{2}{*}{\textbf{PWB6K(n/N)}}  &  \multirow{2}{*}{\textbf{HF-3c(n/N)}}    & \multirow{2}{*}{\textbf{CCSD(T)/CBS}}   & \multirow{2}{*}{\textbf{$\boldsymbol\omega$-PBE}} \\  
   
  & &  \\
\hline
 \ce{NH3}       & 0.06 (4/7) &  0.14  (5/7)       &  	 	3632, 3516, 3562 & 79 \\
  \ce{CH3OH}    &   0.08 (8/8) & 0.13 (6/8)  &  3922, 	4111, 4005 & 119 \\ 
  \ce{HCOOH} & 0.06 (11/13) & 0.17 (10/13) &  & \\  	
   \ce{H2CO} &    0.06 (5/6) & 0.15 (4/6) & 2600, 	1197, 1181 & 338 \\
    HF &  0.04 (3/4) & 0.06 (2/4) & 5956, 	5380, 	4158 & 83\\
  HCl  & 0.07 (6/6) & 0.18 (2/6) & 3445, 2923, 	956 & 146 \\ 
 HCO &  0.05 (3/3)& 0.07 (1/3) &  2224, 	1684 & 56 \\
 HNC & 0.08 (4/5) & 0.30 (3/5) & 4211,3953  & 305 \\
 HCN & 0.06 (4/5) & 0.20 (3/5) \\
 \hline
 Average &   0.06 (48/57)  & 0.15 (36/57) & & 160 \\
 \hline
\end{tabular*}
\end{table}

\noindent Regarding the energy benchmark, Table \ref{tab:bench}, columns 4-5, for both groups the best DFT functional is the $\omega$-PBE \citep{vydrov2006,vydrov2006a,vydrov2007} 
with BSSE and D3BJ 
dispersion corrections, coupled with def2-TZVP basis set. The average mean absolute error (MAE) 
is 37 and 160 K for Group D and H respectively.
Full benchmark results can be found at \url{www.github.com/QCMM/beep}.

{}

\clearpage

\section{Gaussian fitting procedure}\label{sec:boot}
\noindent To fit the BE distribution data with a Gaussian function, we employed a
bootstrap method. We first divide our sample in equally-spaced bins, so that each bin 
contains $N_i$ samples, with a Poisson error $\sqrt{N_i}$. We then produce $10^4$ distributions
analogue to the original data, randomizing the points 
assuming a Gaussian error of
$\sqrt{N_i}$ around the mean $N_i$ and we fit each distribution with
\begin{equation}\label{eq:fit}
 f(x) = a \exp\left(-\frac{(x-\mu)^2}{2\sigma^2}\right)\,,
\end{equation}
where $a$, $\mu$, and $\sigma$ are free parameters. The binned distribution of each parameter 
after the $10^4$ iterations is also a Gaussian, where the average is the value we assume for
the given parameter and the dispersion is the associated error.

\section{Dispersion correction}\label{sec:dispersion}
\noindent The following figure reports the comparison between the histograms of the BE distributions computed in this work, with and without including dispersion correction (D3BJ). 
The low impact of this contribution on the BE  is reflected in a small shift of the distributions for most of the molecules in Group H, lower panel. On the other hand, the D3BJ correction is essential for Group D molecules, upper panel,  as it shifts the BE distributions into the bound regime. 

\begin{figure*}[h]
    \centering
\includegraphics[angle=0, width = 0.9\linewidth]{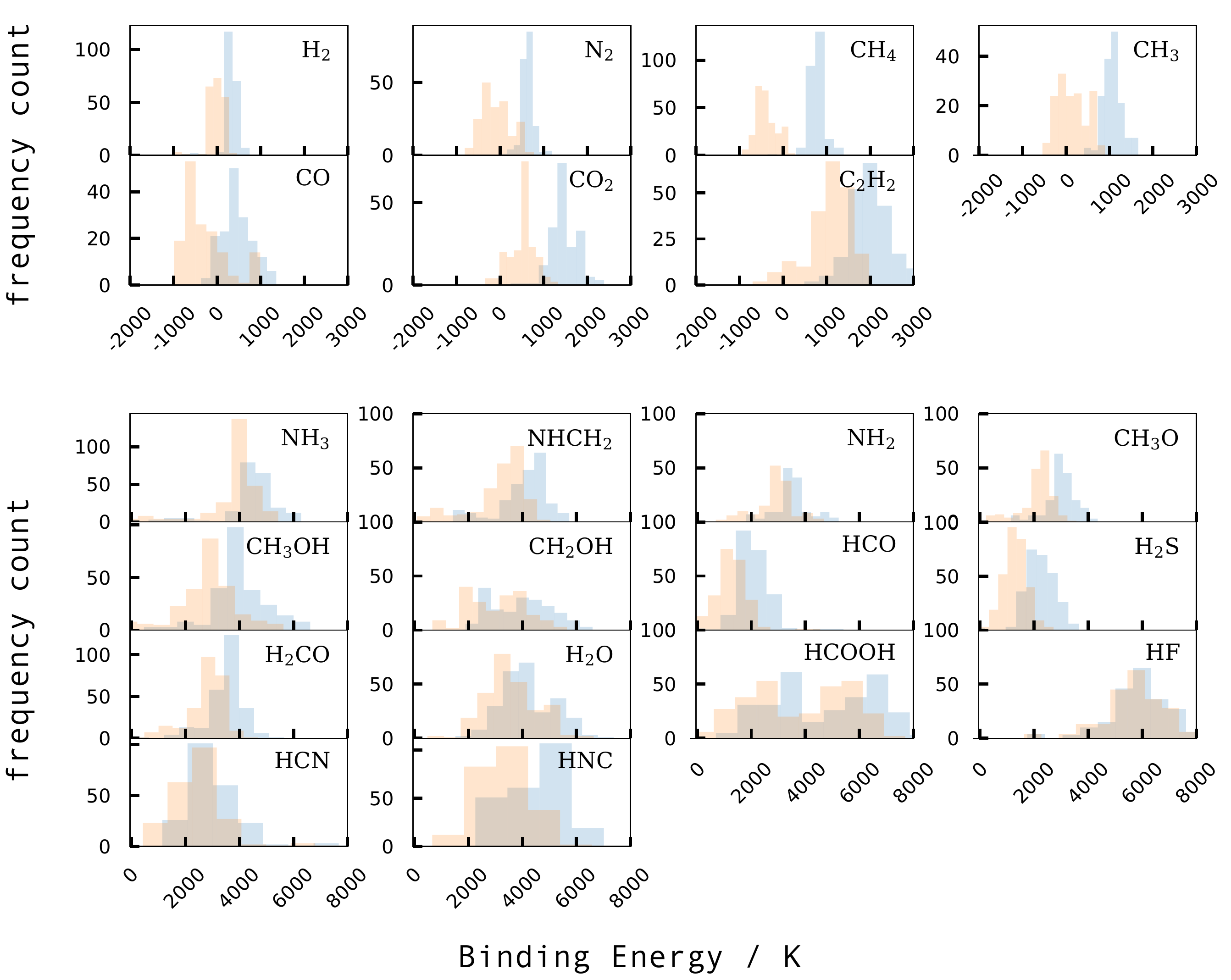}
    
    \caption{Binding energy distributions computed  using the best performing DFT functional from the energy benchmark for each molecule, with (blue) or without (orange) including D3BJ correction. Upper panel: Group D; lower panel: Group H.}
    \label{fig:dispersion}
\end{figure*}{}

\clearpage

\section{Astrophysical framework}\label{sec:framework_astro}
\noindent We assume a protoplanetary disk with a gas and dust temperature radial profile on the midplane $T_d(R) = T(R) = T_0 (R/1\,{\rm au})^{-0.5}$, with $T_0=200$\,K. The $\alpha$-viscosity $\nu(R) = \alpha c_{\rm s}^2(R) \Omega_{\rm K}^{-1}(R)$ depends on the thermal speed of sound $c_{\rm s}(R)=\sqrt{k_{\rm B} T(R) \mu^{-1} m_{\rm p}^{-1}}$, where the mean molecular weight is $\mu=2.34$, and $m_{\rm p}$ is the mass of the proton, and on the Keplerian angular frequency $\Omega_{\rm K}=\sqrt{G M_{*} R^{-3}}$, where $G$ is the gravitational constant, and $M_*=1\,$M$_\odot$ is the mass of the central star. With these definitions, the constants in the main text are
\begin{eqnarray}
    \varphi_1 &=& k_{\rm B} T_0 \sqrt{R_0}\,,\\
    \varphi_2 &=& \frac{\mu m_{\mathrm p}}{\alpha k_{\rm B} T_0} \nu_0 \sqrt{\frac{G M_*}{R_0}}\,,
\end{eqnarray}
where $R_0=1$\,au is the position where $T(R)=T_0$, and $\nu_0=10^{12}$\,s$^{-1}$ is the Debye frequency. 

\clearpage

\end{document}